\documentclass{JHEP3}

\newcommand{\al}{\alpha}
\newcommand{\bara}{\bar{\alpha}}

\newcommand{\tr}{{\mathrm{tr}}}

\newcommand{\one}{\! \hbox{ 1\kern-.8mm l}}
\newcommand{\be}{\begin{equation}}
\newcommand{\ee}{\end{equation}}
\newcommand{\demu}{\partial_{\mu}}

\newcommand{\bear}{\begin{array}{l}}
\newcommand{\eear}{\end{array}}
\newcommand{\ds}{\displaystyle}

\newcommand{\inte}{\! \int \!\!}

\newcommand{\infNdu}{\int \!\! {\cal D}[V] \, {\cal D}[\varphi]
\, {\cal D} [\overline{\varphi}] \prod_{i=1}^{2} {\cal D}[\varphi_i]
\, {\cal D} [\overline{\varphi}_i]  }
\newcommand{\infNdupr}{\int \!\! {\cal D}[V] \, {\cal D}[\varphi]
\, {\cal D} [\overline{\varphi}] \prod_{i=1}^{2} {\cal D}[\varphi'_i]
\, {\cal D} [\overline{\varphi}'_i]  }
\newcommand{\mdm}{M \partial_{M}}

\newcommand{\ie}{{\it i.e.}\ }

\newcommand{\aka}{{\it a.k.a.}\ }

\newcommand{\ug}{\!=\!}
\newcommand{\um}{\frac{1}{2}}
\newcommand{\N}{ {\cal N}}
\newcommand{\nn}{{\rm I}\hspace{-.18em}{\rm N}}
\newcommand{\mt}{ \widetilde{M}}
\newcommand{\st}{ \tilde{S}}
\newcommand{\nfo}{ ${\cal N}\!=\! 4$}
\newcommand{\ntwo}{ ${\cal N}\!=\! 2$}
\newcommand{\none}{ ${\cal N}\!=\! 1$}
\def\eq#1{eq.~(\ref{#1})}
\def\ceq#1{Eq.~(\ref{#1})}
\def\dphi#1#2{\frac{\delta #1}{\delta \varphi_{#2}} }
\def\dphip#1#2{\frac{\delta #1}{\delta \varphi'_{#2}} }

\def\dphib#1#2{\frac{\delta #1}{\delta \overline{\varphi}_{#2}} }
\def\ddphi#1#2#3{\frac{\delta^2 #1}{\delta \varphi_{#2} \, \delta
\varphi_{#3} } }

\def\d#1{\delta_{#1}}
\def \fume#1#2{\prod_{i=1}^{#1} {\cal D}[\varphi^{#2}_i]
\, {\cal D} [\overline{\varphi}^{#2}_i]}

\title{Exact renormalization group equation in presence of rescaling
anomaly II -\\The local potential approximation} 

\author{S. Arnone${}^a$, D. Francia${}^b$
and K. Yoshida${}^b$\\ 
${}^a$Department of Physics and Astronomy,
University of Southampton\\
\hspace{.045em} Highfield, Southampton SO17 1BJ, U.K.\\

${}^b$Dipartimento di Fisica,
Universit\`a degli Studi di Roma ``La Sapienza''\\
\hspace{.045em} P.le Aldo Moro, 2 - 00185 Roma, Italia
and
I.N.F.N., Sezione di Roma I\\
\ \\
\ \\
E-mails: \email{sa@hep.phys.soton.ac.uk, dario.francia@roma1.infn.it, 
kensuke.yoshida@roma1.infn.it}}  

\maketitle

\preprint{SHEP 01-28}
\abstract{Exact renormalization group techniques are applied to mass
deformed\nfo \  supersymmetric Yang--Mills theory, viewed as a regularised\ntwo \
model. The solution of the flow equation, in the local potential
approximation, reproduces the one-loop (perturbatively exact) expression for
the effective action of\ntwo \ supersymmetric Yang--Mills theory, when the
regularising mass, $M$, reaches the value of the dynamical cutoff
$\Lambda$. One speculates about the way in which further non-perturbative
contributions (instanton effects) may be accounted for.}

\begin{document}
\section{Introduction}

In a previous paper~\cite{noi}, we have studied the proposals by
Arkani-Hamed and Murayama~\cite{mu1,mu2} concerning the renormalization
group (RG) invariance of exact results in supersymmetric (SUSY) gauge field
theories. Our main tool was the method of the exact renormalization group
(ERG) equation~\cite{noi}, inspired by the decimation method by
Wilson~\cite{Wil} and adapted by Polchinski and Gallavotti~\cite{Po} to the
continuum field theory case.

The main results in~\cite{noi} are:

-- the dominant part of the low energy effective (\aka ``Wilsonian'' )
  action   
  which is responsible for the Novikov-Shifman-Vainshtein-Zakharov (NSVZ)
  exact expression for the beta function~\cite{NSVZ} is the anomalous term due
  to the rescaling (\aka Konishi) anomaly~\cite{Kon};

-- after subtracting such an anomalous term, the remainder of the effective
  action satisfies the ERG equation, or Polchinski's equation, with respect
  to the variation of the regularising mass, $M$.

After these results, as a further test of our method, we would like to see
whether it can be used to get further useful information about the low
energy interactions of SUSY gauge field theories, beyond the NSVZ formula.\\
In particular, it should be possible, at least in principle, to compute the
low energy effective action for the \ntwo \ SUSY Yang--Mills theory (SYM), as
regularised as the ultraviolet (UV) finite mass deformed\nfo \
SYM~\cite{kovacs}. 

In a subsequent note~\cite{procromayosh}, however, we adopted the idea that
the detailed structure of low energy effective action could be obtained by
generalising the ``anomaly analysis'' in~\cite{noi,Po} rather than seeking
the solution of the ERG equation we derived in~\cite{noi}. 
As a matter
of fact, in~\cite{procromayosh} it has been tacitly assumed that the
``non-anomalous'' part of the effective action becomes irrelevant once the
generalised Konishi anomaly has been introduced. 

The expression of the low energy effective action (the vector field
kinematical term) of\ntwo \ SYM given in~\cite{procromayosh}, although
similar in functional form to the standard perturbately exact one-loop
result, has the wrong dependence on the relevant mass parameters ($M$ and
$M_0$, the latter being the starting point of the flow), so that one cannot
reproduce the correct exclusive dependence on the dynamical cutoff
$\Lambda$.

In the present note, we would like to pursue the original idea
in~\cite{noi} 
and look back at the ERG equation for the non-anomalous part of
the effective action. We will try and solve that equation in the simplest
standard approximation, \ie the local potential approximation (LPA).

It might be expected that, within such approximation, one would essentially
end up with the same result as in~\cite{procromayosh}. Instead, the
effective action we computed shows similar functional dependence on the
chiral superfield $\varphi$ but different dependence on the mass
parameters. This change brings the new result into complete agreement with the
standard perturbately exact result for\ntwo \ SYM in weak coupling regime.

In~\cite{noi}, Polchinski's original derivation of the ERG equation has
been closely followed. With this method one sees the appearance of a
singular functional derivative term which is then identified with the
contribution of the rescaling anomaly. Even though this passage can be made
more convincing by using some pre-regularisation, it would be nice to avoid
any reference to such a singular term. This can be achieved by applying the
method proposed by C. Becchi~\cite{Bec}. 
In the standard RG approach, this method consists in 
damping high modes by a suitable field redefinition, $\phi_p \rightarrow
K_\Lambda \phi_p$, in the interaction part of the action only. Thus, up to
a vacuum energy term, it is equivalent to modifying the free propagator,
$D(p) \rightarrow D(p) K_\Lambda^{-1}$. $K_\Lambda \equiv
K \left( {p^2 \over \Lambda^2} \right)$ is a smooth ultraviolet cutoff
profile, rapidly vanishing at infinity.

The paper is organised as follows: in sec.\ 2 we introduce the mass
deformed \nfo \ SYM theory and re-derive 
the ERG equation in presence of the rescaling anomaly by generalising
Becchi's transformation. Sec.\ 3 is devoted to the study of the flow
equation in the LPA: we start off with the $SU(2)$ case and, then,
generalise the results to $SU(N)$. In sec.\ 4 we summarise and draw our
conclusions.  

\section{Exact renormalization group equation for SUSY gauge theories}

As in~\cite{noi}, we consider the mass deformed version of\nfo \  $SU(N)$ SYM
in four space-time dimensions, corresponding to the so-called 
${\cal N} \ug 1^*$ or ${\cal N} \ug 2^*$ models. 

The classical action is given by $S^* = S_{{\cal N}=4} +$ 
mass terms and 
reads (written in terms of \none
\ superfields and in the ``holomorphic'' representation)
\be \label{esseN4}
\bear
{\ds S^* (V,\varphi_i,\overline{\varphi}_i; g_0) = \frac{1}{16} \inte d^4 x \, d^2
\theta \frac{1}{g_0^2} W^a_{\alpha} W^{a \alpha} +
\inte dx \, d^4 \theta  \,{\Re}e  \left( \frac{2}{g_0^2}\right) 
N \sum_{i=1}^3 \overline{\varphi}_i e^V \varphi_i 
+}\nonumber\\[0.3cm]
{\ds + \inte d^4 x \, d^2 \theta \, {\Re}e  \left( \frac{\sqrt{2}}{g_0^2} 
\right) i N f_{abc} \, \varphi^i_a \, \varphi^j_b \, \varphi^k_c
\, \frac{\epsilon^{ijk}}{3!} + \frac{1}{2} \inte d^4 x \, d^2
\theta \sum_{i=1}^{\mu} M_i^0 \varphi_i^2 + h.c., }\\
\eear
\ee 
all the relevant fields transforming as the adjoint representation of the
gauge group.

In order to break supersymmetry down to\ntwo \ or\none, one gives mass to 
some of the chiral superfields in the\nfo \ supermultiplet. These masses
are supposed be large compared to the range of momenta one is interested
in. 
 In what follows we will consider the case in which two chiral superfields
are given the same mass, \ie $\mu \ug 2$ and $M_1^0 \ug M_2^0 \ug M_0$, as it
preserves\ntwo \ SUSY.   

This model, just as the original\nfo \ SYM theory without chiral
mass terms, is believed to be finite. 
It has been shown that, at the perturbative level, all the 
UV divergences cancel out~\cite{kovacs}.

The generating functional takes the form
\be \label{zeta}
\bear
{\ds Z_{M_0} = \infNdu \exp i \bigg[ S_{{\cal N}=4} (V,\varphi,
\overline{\varphi},\varphi_i,\overline{\varphi}_i;
g_0) +}\nonumber\\[0.3cm]
{\ds + \frac{M_0}{2} \int \sum_{i=1}^{2} \varphi_i^2 + h.c. + 
  
\int \sum_{i=1}^{2} J_i \, \varphi_i + h.c. +
\inte J_{\varphi} \, \varphi + \inte 
J_{\overline{\varphi}} \, \overline{\varphi} + \inte J_V V \bigg],}
\eear
\ee
where $\varphi$ is the massless chiral superfield before denoted by
$\varphi_3$ and the integration over the chiral or full superspace has not been
written explicitly.\footnote{In order to simplify notation, we will dispense
with the integration measure in what follows.} 
The field set $\left(V,\varphi,\overline{\varphi}\right)$ is
often referred to as 
the\ntwo \ vector multiplet. 

The ERG method consists in varying the regularising mass $M_0$ to $M \leq
M_0$ and compensating for such change  by replacing the bare action
by the Wilsonian effective action $S_M$, \ie
\be \label{zetam0eqzetam}
\bear
{\ds Z_{M_0} = Z_M \equiv \infNdu \, \exp i \bigg[ S_M (V,\varphi,\overline{\varphi},\varphi_i,\overline{\varphi}_i) + 
 }\nonumber\\[0.25cm]   
{\ds + \frac{M}{2} \int \sum_{i=1}^{2} \varphi_i^2
+ \int \sum_{i=1}^{2} f(M) \, J_i \, \varphi_i +  
\frac{1}{2} \int \sum_{i=1}^{2} g(M) \, J_i^2 + \int 
J_{\varphi} \, \varphi + h.c. + \int J_V V \bigg]. }
\eear  
\ee

Demanding the generating functional be invariant under the continuous change 
 in the parameter $M$ one gets the equation obeyed by the effective action
 $S_M$. 

Following the method in~\cite{Bec}, we implement Becchi's
 transformation by rescaling auxiliary chiral superfields,
$$
\varphi_i = \sqrt{{M_0 \over M}} \, \varphi'_i, \qquad \qquad
 \overline{\varphi}_i = \sqrt{{M_0 \over M}} \, \overline{\varphi}'_i,
 \quad \qquad i=1,2. 
$$
In this way, the dependence upon $M$ is transferred from the mass term to
 the effective action itself. Moreover, the functional measure for the
 relevant fields is known to acquire the non-trivial Jacobian
 determinant~\cite{Kon}
$$
\fume{\mu}{} = \fume{\mu}{'} \exp \left[ -{i\over 2} \inte d^4 x \, d^2
 \theta \, {\mu N 
 \over 64 \pi^2} 
 \log \left( {M \over M_0} \right) \, W^a_{\alpha} W^{a \alpha} +
 h.c. \right],   
$$
where $\mu \ug 2$ in the present case and the factor $-\um$ is due to $\log
 \sqrt{{M_0 \over M}} = -\um \log \left( {M \over M_0} \right)$.

Thus $Z_M$ can be rewritten as 
\be \label{zetam}
\bear
{\ds Z_M = \infNdupr \, \exp i \bigg[ S_M }
\left(V,\varphi,\overline{\varphi},\sqrt{{M_0 \over M}} \varphi'_i, \sqrt{{M_0
\over M}} \overline{\varphi}'_i \right) 
+ \nonumber\\[0.25cm]   
{\ds - {N \over 64 \pi^2} \inte \log \left( {M \over M_0} \right) \,
W^a_{\alpha} W^{a \alpha} + h.c. +
\frac{M_0}{2} \inte \sum_{i=1}^{2} \varphi_i^{'2}
+ \inte \sum_{i=1}^{2} f(M) }\sqrt{{M_0 \over M}} {\ds \, J_i \, \varphi'_i
+ }\nonumber\\[0.25cm] 
{\ds +\frac{1}{2} \inte \sum_{i=1}^{2} g(M) \, J_i^2 + h.c. + \inte 
J_{\varphi} \, \varphi + \inte 
J_{\overline{\varphi}} \, \overline{\varphi} + \int J_V V \bigg]. }
\eear  
\ee

As pointed out before, the equation for $S_M$ follows from the invariance
of the generating functional, $Z_{M_0} = Z_M$, or equivalently $\mdm Z_M = 0$.

\be \label{chain1}
\bear
{\ds 
\mdm Z_M = i \infNdupr \, \bigg[ - {N \over 64 \pi^2} \inte W^a_{\alpha}
W^{a \alpha} +
h.c. +}\nonumber\\[0.25cm] 
{\ds + \mdm S_M \Big|_{\varphi_i,
\overline{\varphi}_i} +\mdm } \sqrt{M_0 \over M} {\ds \inte
\varphi'_i \, \dphi{S_M}{i} + \mdm} \sqrt{M_0 \over M} {\ds \inte
\overline{\varphi}'_i \, \dphib{S_M}{i} + }\nonumber\\[0.25cm]
{\ds + \mdm \inte f(M)}\sqrt{{M_0 \over M}}
{\ds \, J_i \, \varphi'_i    
+ \frac{1}{2} \mdm \inte g(M) \, J_i^2 + h.c. \bigg] \exp i S_{tot},  }
\eear
\ee
where $S_{tot}$ stands for the argument of the exponential in \eq{zetam}
and the sum upon $i$ has been left out.

Noting that the equation of motion
$\varphi'_i = {1 \over M_0} \left( \dphip{S_{tot}}{i} - \dphip{S_M}{i} -
f(M) \sqrt{M_0 \over M} J_i
\right)$ is valid within the functional integral, 
the r.h.s. of \eq{chain1} can be recast as
\be
\bear \label{chain2}
{\ds i \infNdupr \bigg[ - {N \over 64 \pi^2} \inte W^a_{\alpha} W^{a \alpha} + h.c. + \mdm S_M \Big|_{\varphi_i,
\overline{\varphi}_i} + }\nonumber\\[0.35cm]
{\ds - {1 \over 2 M} \int \dphi{S_M}{i} 
\bigg( \dphip{S_{tot}}{i} - \dphip{S_M}{i} - f(M)}\sqrt{M_0 \over M} {\ds J_i
\bigg) + \mdm \inte f(M)}\sqrt{{M_0 \over M}}
{\ds \, J_i \, \varphi'_i + }\nonumber\\[0.35cm]
{\ds + \frac{1}{2} \mdm \inte g(M) \, J_i^2   
 \bigg] \exp i S_{tot}.  }
\eear
\ee

Integrating by parts, so that $\dphi{S_M}{ia} \dphi{S_{tot}}{ia}$
becomes $i \ddphi{S_M}{ia}{ia}$, and making again use of the above 
equation of motion , \eq{chain2} becomes 

\be
\bear \label{chain3}
{\ds i \infNdupr \bigg[ \mdm S_M \Big|_{\varphi_i,
\overline{\varphi}_i} - {1 \over 2 M} \inte  
\bigg( i \ddphi{S_M}{i}{i} - \dphi{S_M}{i} \bigg) +
}\nonumber\\[0.35cm]
{\ds - {1 \over 2 M} \inte  
\left( \sqrt{M_0 M} \varphi'_i + f(M) J_i \right) f(M) J_i + \mdm \inte
f(M)}\sqrt{{M_0 \over M}} 
{\ds \, J_i \, \varphi'_i + }\nonumber\\[0.35cm]
{\ds +\frac{1}{2} \mdm \inte g(M) \, J_i^2   
 - {N \over 64 \pi^2} \inte W^a_{\alpha} W^{a \alpha} + h.c. \bigg] \exp i S_{tot}.  }
\eear
\ee
The equations for $f(M), g(M)$ simply follow from the inspection of
\eq{chain3}, 
$$
\mdm f(M) = f(M), \qquad \qquad \mdm g(M) = \frac{1}{M}
f^2(M).
$$
The solutions to the above equations with the proper boundary conditions
-- $f(M_0)=1$ and $g(M_0)=0$ -- are 
$f(M)={M \over M_0}$ and $g(M)={M \over M_0^2} - {1\over M_0}$. 

Hence the generating functional is indeed invariant under the RG
transformation if the effective action satisfies the anomalous Polchinski's
equation
$$
\mdm S_M = \frac{1}{2 M} \inte \left[ i \ddphi{S_M}{i}{i} 
- \dphi{S_M}{i} \dphi{S_M}{i} \right] + h.c. + {N \over 64 \pi^2} \int
W^a_{\alpha} W^{a \alpha} + h.c.
$$
As in~\cite{noi}, the above equation can be also recast as
\be \label{main}
\mdm \tilde{S}_M = \frac{1}{2 M} \int \left[ i \ddphi{\tilde{S}_M}{i}{i} 
- \dphi{\tilde{S}_M}{i} \dphi{\tilde{S}_M}{i} \right] + h.c.,
\ee
where $\tilde{S}_M = S_M  - {N \over 64 \pi^2} \int \log \left( {M\over M_0}
\right) \, W^a_{\alpha} W^{a \alpha} + h.c.$, as the contribution of the anomaly, $\tilde{S}_M -
S_M$, does not 
depend upon the auxiliary massive fields. 

\ceq{main} is the main result in~\cite{noi}. As stated in the introduction,
in the present derivation singular terms such as $\sum_i \dphi{\varphi_i}{i}$  
do not appear at all.\footnote{K.Y. is grateful to T.Eguchi for pointing
out this problem in [1].} 

\section{The local potential approximation to Polchinski's equation}

As outlined in the introduction, we would like to find the solution of
\eq{main} within some approximation scheme, different from the standard
perturbative loop expansion.
For the relevant initial condition at $M\ug M_0$, we will try and determine
$\tilde{S}_M$ at the lowest order approximation in the derivative
expansion, \ie the local potential approximation. 

In order to carry out the LPA consistently, one usually first Legendre
transforms the Wilsonian action $S_{\Lambda}$ ($\Lambda$ being the cutoff, in
the present case the regularising mass parameter) to the
one-particle-irreducible action $\Gamma_{\Lambda \Lambda_0}$~\cite{Wet}, with
$\Lambda$  now acting as an infra-red cutoff, and then applies the
Coleman-Weinberg expansion
$$
\Gamma_{\Lambda \Lambda_0} (\phi) = \inte d^4 x \, V_{\Lambda \Lambda_0}
(\phi) + \inte d^4 x \, Z_{\Lambda \Lambda_0} (\phi) (\demu \phi)^2 + \cdots,
$$      
with the ellipsis standing for higher derivative terms. Substituting
$\Gamma_{\Lambda \Lambda_0} (\phi)$ back into the flow equation, the various
coefficients, $V_{\Lambda \Lambda_0} (\phi),  Z_{\Lambda \Lambda_0} (\phi),
\ldots,$ can be consistently determined~\cite{Wet}.  

In the present note, since we are interested in the lowest order local
potential term in a range of momenta much lower than the regulating mass, 
we will try the following {\it ansatz} (for the details, please refer to
T.R. Morris in~\cite{Wet})
$$
\st_M (V,\varphi,
\overline{\varphi},\varphi_i,\overline{\varphi}_i) \simeq \inte d^4 x \,
d^2 \theta \, V_M (V,\varphi, \varphi_i) + h.c., 
$$
where $V_M$ is assumed to be an holomorphic function of the chiral fields.
The initial condition, $V_{M_0}$, can be read off from $S_{\N = 4}$ 
\be \label{lpa}
V_{M_0} (V,\varphi, \varphi_i) = \int i \alpha \, f_{abc} \,\varphi^a
\, \varphi^b_1 \, \varphi^c_2 = \int \alpha \left( -\hat{F} \cdot
\vec{\varphi} \right)_{ab} 
\varphi^a_1 \varphi^b_2,
\ee
where $\alpha = \sqrt{2} N \, {\Re}e  \left( \frac{1}{g_0^2} 
\right)$ and $\hat{F}$ represents the set of hermitian generators of $SU(N)$
in the adjoint representation.

As was shown in~\cite{noi}, Polchinski's equation can be formally solved 
by the integral form
\be \label{intfor}
\bear
{\ds \exp i \st_M \left(V,\varphi, \overline{\varphi}, \varphi_i, \overline{\varphi}_i \right) = 
\Bigg \{ \inte \prod_{i=1}^2 {\cal D}[\varphi'_{i}] {\cal D}[\overline{\varphi}'_i] \exp i 
\bigg[ \st_{M_0} \left(V, \varphi, \overline{\varphi}, \varphi'_i,
\overline{\varphi}'_i \right) + }\nonumber\\[0.35 cm]
{\ds + \frac{\mt}{2} \inte \sum_{i=1}^2 \left( \varphi'_{i} - \varphi_i
\right)^2 + h.c. 
\bigg ] \Bigg \} \Bigg \{ \inte \prod_{i=1}^2 {\cal D}[\varphi'_{i}] 
{\cal D}[\overline{\varphi}'_i] 
\exp i \bigg[ \frac{\tilde{M}}{2} \inte \sum_{i=1}^2 \varphi_{i}^{'2} +
h.c. \bigg] \Bigg \}^{-1}, }\\
\eear
\ee
where the ``reduced'' mass $\mt$ is defined by $\mt^{-1} \doteq 
M_0^{-1} - M^{-1}$.
The expression (\ref{intfor}) is formal in that the convergence of the
r.h.s. is not well established.

In the LPA, $\st_{M_0}$ is to be replaced by \eq{lpa}, which is a quadratic
form in the integration variables, $\varphi_i$,
\be \label{A}
\st_{M_0}\left(V, \varphi, \overline{\varphi}, \varphi'_i,
\overline{\varphi}'_i \right) = \um \int \varphi'_{ia} \, \hat{A}_{ia,jb} 
(\varphi) \, \varphi'_{jb} + h.c.,
\ee
with $\hat{A}_{ia,jb} (\varphi) \doteq - \alpha \, \epsilon_{ij} \left(
\hat{F} \cdot \vec{\varphi} \right)_{ab}$.

Hence the Gaussian integral
\be \label{intfor2}
\bear
{\ds Z' = 
\Bigg \{ \inte \prod_{i=1}^2 {\cal D}[\varphi'_{i}] {\cal D}[\overline{\varphi}'_{i}] \exp i 
\bigg[ \um \int \varphi'{}^T \, \hat{A} 
(\varphi) \, \varphi' + \frac{\mt}{2} \inte \sum_{i=1}^2 \left(
\varphi'_{ia} - \varphi_{ia} \right)^2 +  h.c. \bigg ] \Bigg \} \times
}\nonumber\\[0.35cm]  
{\ds \qquad \,\, \times \Bigg \{ \inte \prod_{i=1}^2 {\cal
D}[\varphi'_{i}] {\cal D}[\overline{\varphi}'_i] 
\exp i \bigg[ \frac{\tilde{M}}{2} \inte \sum_{i=1}^2 \varphi_{ia}^{'2} +
h.c. \bigg] \Bigg \}^{-1} }\\
\eear
\ee 
is to be evaluated. 

Transforming the r.h.s. of \eq{intfor2} by quadrature yields 
\be \label{intfor3}
\bear
{\ds Z' = \exp i{\mt \over 2} \bigg[ \int \varphi_{ia} \left(
\! \one - \mt \left( \hat{A} + \mt \! \one \right)^{-1} \right)_{ia,jb} \,
\varphi_{jb} + h.c. \bigg] \times }\\[0.35cm]
{\ds \times \Bigg \{ \inte \prod_{i=1}^2 {\cal D}[\varphi'_{i}] {\cal
D}[\overline{\varphi}'_{i}] \exp {i \over 2} \bigg[
 \int \varphi'_{ia} \left( \hat{A} + \mt \one \right)_{ia,jb}  
 \varphi'_{jb} + h.c. \bigg ] \Bigg \} \times
}\nonumber\\[0.35cm] 
{\ds \times \Bigg \{ \inte \prod_{i=1}^2 {\cal D}[\varphi'_{i}] 
{\cal D}[\overline{\varphi}'_i] 
\exp {i \over 2} \bigg[ \inte \sum_{i=1}^2 \mt \varphi_{ia}^{'2} +
h.c. \bigg] \Bigg \}^{-1}. }
\eear
\ee 

The ratio of the Gaussian integrals in \eq{intfor3} can be simply computed by
applying the matrix rescaling anomaly~\cite{Kon}.
Writing 
$$
\varphi'_{ia} = \left[ \mt \left( \hat{A} + \mt \! \one \right)^{-1}
\right]^{\um}_{ia,jb} \varphi^{''}_{jb} = \left[ \! \one - \left( 1
- {M\over M_0} \right) {\hat{A} \over M} \ds \right]^{-\um}_{ia,jb}
\varphi^{''}_{jb}, 
$$
one has 
$$
\prod {\cal D}[\varphi'_{i}] = \prod {\cal D}[\varphi^{''}_{i}] \exp i
\left[ {1\over 128 \pi^2} \int \tr \left\{ \log \left(\! \one - \left( 1
- {M\over M_0} \right) {\hat{A} \over M} \right) W^2 \right\} + h.c. \right].
$$
The kinematical term for the vector superfield in $S_M$ takes the form
\be \label{kin}
{1\over 16} \int \left[ \left( {1\over g_0^2} + {2 N\over 8 \pi^2} \log
{M\over M_0} \right) W^a_{\alpha} W^{a \alpha} + {1\over 8\pi^2} \tr \left\{ \log \left(\! \one - \left( 1
- {M\over M_0} \right) {\hat{A} \over M} \right) W^2 \right\} \right] +
h.c.
\ee 
Since we are interested in the low energy dynamics only, we will restrict
ourselves to massless configurations. In the $SU(2)$ case, this amounts to
choosing the particular configuration $\vec{V}=(0,0,V)$ and $\vec{\varphi}=
(0,0,\varphi)$. For such $SU(2)$ configuration, \eq{kin} becomes
\be \label{kin2}
{1\over 16} \int \left[ \left( {1\over g_0^2} + {1\over 2 \pi^2} \log
{M\over M_0} \right) W^a_{\alpha} W^{a \alpha} + {1\over 4\pi^2} \log \left(1+ \alpha^2 \left( 1
- {M\over M_0} \right)^2 {\varphi^2 \over M^2} \right) W^a_{\alpha} W^{a
\alpha} \right] +
h.c.
\ee
Hence, the effective action $S_M$ is approximately given by
\be \label{smlpa}
\bear
{\ds S_M \left(V,\varphi, \overline{\varphi}, \varphi_i, \overline{\varphi}_i
\right) = {1\over 16} \int \left[ \left( {1\over g_0^2} + {1\over 2 \pi^2} \log
{M\over M_0} \right)  W^a_{\alpha} W^{a \alpha} + {1\over 4\pi^2} \log \left(1+ \alpha^2 \left( 1
- {M\over M_0} \right)^2 \right. \right. \!\! \times }\\[0.25cm]
{\ds \times \left. \left.  {\varphi^2 \over M^2} \right) W^a_{\alpha} W^{a
\alpha} \right] +
h.c. + {\mt \over 2} \bigg[ \int \varphi_{ia} \left(
\! \one - \left(\! \one - \left( 1
- {M\over M_0} \right) {\hat{A} \over M} \right)^{-1} \right)_{ia,jb} \,
\varphi_{jb} + h.c. \bigg]. }
\eear
\ee
Let us now specialise to the weak coupling regime, \ie ${1\over g_0^2}>>1$.
It is convenient to introduce the dynamically generated scale $\Lambda$,
defined as 
$\Lambda = M_0 \exp -{2 \pi^2\over g_0^2}$, such that ${1\over
g^2(\Lambda)} = {1\over g_0^2} + {1\over 2\pi^2} \log {\Lambda \over M_0} = 
0$, \ie the running coupling diverges at $M=\Lambda$.

When ${1\over g_0^2}$ is much bigger than $1$, $\Lambda$ is much smaller
than $M_0$. Moreover, if one replaces the field variable $\varphi$ with its
vacuum expectation value (vev), then one can also assume $|\varphi| >>
\Lambda$. In this regime the quantum partition function of the mass
deformed theory is given by
$$
Z_\Lambda \simeq \infNdu \exp i \bigg[ S_{\Lambda} + {\Lambda \over 2} \inte
\sum_{i=1}^2 \varphi_i^2 + h.c. + \cdots \bigg],
$$
and the form of $S_\Lambda$ can be read off from \eq{smlpa}.\\ 
The term
sandwiched between the massive fields reduces to 
$$
\one - \left(\! \one - \left( 1
- {M\over M_0} \right) {\hat{A} \over M} \right)^{-1} 
\simeq \! \one - \left(\! \one - {\hat{A} \over M} \right)^{-1} 
\simeq -{\hat{A} \over
\Lambda} \left(\! \one - {\hat{A} \over
\Lambda} \right)^{-1} \simeq \! \one,
$$
so that the last term in \eq{smlpa} is given by $ - {\Lambda \over 2} \inte
\sum_{i=1}^2 \varphi_i^2 + h.c.$

Thus, for large $|{\varphi \over \Lambda}|$ the total action $S_{tot}$ can
be written as
\be \label{correct}
S_{tot} = S_{\Lambda} + {\Lambda \over 2} \inte
\sum_{i=1}^2 \varphi_i^2 + h.c. + \cdots \simeq {1\over 64 \pi^2} \inte
\log \left( {\varphi^2\over \Lambda^2} \right)  W^a_{\alpha} W^{a \alpha} + h.c. + {\rm source \ terms}.
\ee

The last expression agrees with the standard ``perturbatively exact''result
for \ntwo \, $SU(2)$ SYM.

In~\cite{procromayosh}, we have tried to interpret all the relevant part of
the effective action as resulting from the effect of the Konishi anomaly or its
generalisation. Compared with the present result, eqs (\ref{kin2}), (
\ref{smlpa}), 
the only difference is the dependence  on the ratio ${\varphi \over
M}$. As a matter of fact, in~\cite{procromayosh} $\log (1+ \alpha^2
( 1 - {M\over M_0} )^2 {\varphi^2 \over M^2} )$ is
replaced by $\log (1+ \alpha^2 ( 1 - {M\over M_0} )^2
{\varphi^2 \over M_0^2} )$. 
Although it may not seem so, it makes all the difference, as the latter
expression cannot be recast into the form \eq{correct}, which correctly
depends on the dynamical cutoff $\Lambda$ only. Note that if one naively
integrates out the massive fields in the action, \eq{esseN4}, neglecting the
kinematical term, then one ends up with the corresponding expression $\log
(1+ \alpha^2 {\varphi^2 \over M_0^2})$.
    
\subsection{Generalisation to $SU(N)$}

The explicit results given for $SU(2)$ can be generalised to $SU(N)$ if the
trace of the logarithmic term in \eq{kin} is computed for the
corresponding diagonal (massless) configurations. 
We will discuss the $SU(3)$ case first and, then, generalise the method
by making use of the root equation in the Cartan-Weyl basis of the gauge
group.

In the diagonal configuration, $\vec{\varphi} \cdot \hat{F} = F_3
\varphi_3 + F_8 \varphi_8$, \ie $\varphi_a$ is different from 0 for $a=3,8$
only (Cartan sub-algebra).
 
It is more convenient to re-express this quantity in terms of the
fundamental representation, with redundant field variables, 
$$
\sum_{a=1}^8 \varphi_a \lambda_a = \mathrm{diag} \left( \hat{\varphi}_1,
\hat{\varphi}_2, \hat{\varphi}_3 \right), \qquad \sum_i \hat{\varphi}_i =
0,
$$
with the $\lambda$'s being the
$SU(3)$ generators in the fundamental representation, normalised as $\tr
\lambda_a \lambda_b = \um \d{ab}$.

With this choice of gauge one has
\be \label{gaugechoice}
\hat{\varphi}_1 = {\varphi_3 \over 2} + {\varphi_8 \over 2 \sqrt{3}},
\qquad \hat{\varphi}_2 = -{\varphi_3 \over 2} + {\varphi_8 \over 2 \sqrt{3}},
\qquad \hat{\varphi}_3 = - {\varphi_8 \over \sqrt{3}}
\ee
and
\begin{eqnarray*}
\vec{\varphi} \cdot \! \hat{F} =
\left( \, 
\begin{tabular}{rcl|c|rcl|rcl|c}
&&&&&&&&&& \\
&$\varphi_3 \, \sigma_2$& &&&&&&&&\\ 
&&&&&&&&&& \\ \hline 
&&&$0$&&&&&&& \\ \hline
&&&&&&&&&&\\
&&&&&$\left({\varphi_3\over 2} + {\sqrt{3} \over 2} \varphi_8 \right) \sigma_2$
&&&&&\\
&&&&&&&&&&\\ \hline
&&&&&&&&&&\\
&&&&&&&&$\left(-{\varphi_3\over 2} + {\sqrt{3} \over 2} \varphi_8 \right)
\sigma_2$&&\\
&&&&&&&&&&\\ \hline
&&&&&&&&&&$0$\\
\end{tabular}
\, \right), \\
\underbrace{\hspace{5.3em}}_{1,2}  \,\, \underbrace{\hspace{.4em}}_{3} \,\,\,
\underbrace{\hspace{8.9em}}_{4,5} \,\,\,\,\, \underbrace{\hspace{9.9em}}_{6,7}
\,\,\, \underbrace{\hspace{.5em}}_{8} \hspace{1.3em}\\
\end{eqnarray*}
with all other elements vanishing.

All the non-trivial coefficients in the above table can be written in terms
of $\hat{\varphi}_i - \hat{\varphi}_j$, namely
$$
\Big(a_1, a_2, a_3\Big) \doteq \Big( \varphi_3, {\varphi_3\over 2} + {\sqrt{3}
\over 2} \varphi_8,  -{\varphi_3\over 2} + {\sqrt{3} \over 2} \varphi_8
\Big) = \Big(\hat{\varphi}_1 - \hat{\varphi}_2,\hat{\varphi}_2 - \hat{\varphi}_3,
\hat{\varphi}_1 - \hat{\varphi}_3 \Big).
$$

Thus, the logarithm in \eq{kin2} can be recast as $\log (\one +
\chi \, \sigma_2 \otimes \varepsilon)$ for some constant $\chi$, and can be
evaluated by using  
\be \label{logexp}
\log \left( \one + \chi \, \sigma_2 \otimes \varepsilon \right) = {1\over 2i}
\log \left( \frac{1 + i \chi}{1 - i \chi} \right) \sigma_2 \otimes
\varepsilon + \um \log \left( 1 + \chi^2 \right) \one.
\ee

Repeating the same procedure\footnote{Again only the non-vanishing elements
will be written down.} for $W^2$, 
\begin{eqnarray*}
W^2_{adj} =
\left( \, 
\begin{tabular}{rcl|c|rcl|rcl|c}
&&&&&&&&&& \\
&$W_3^2$& &&&&&&&&\\ 
&&&&&&&&&& \\ \hline 
&&&$0$&&&&&&& \\ \hline
&&&&&&&&&&\\
&&&&&$\left({W_3\over 2} + {\sqrt{3} \over 2} W_8 \right)^2 $
&&&&&\\
&&&&&&&&&&\\ \hline
&&&&&&&&&&\\
&&&&&&&&$\left(-{W_3\over 2} + {\sqrt{3} \over 2} W_8 \right)^2$&&\\
&&&&&&&&&&\\ \hline
&&&&&&&&&&$0$\\
\end{tabular}
\, \right),\\
\end{eqnarray*}
we can
now take the trace over colour indices to get
$$
\bear
{\ds \tr \log \left(\one + \alpha 
\left(1-{M\over M_0}\right)  {{\vec{\varphi} \cdot \! \hat{F} } \over M}
\otimes \varepsilon \right) W^2 = }\nonumber\\
{\ds = \sum_{i>j} \left( \hat{W}_i - \hat{W}_j \right)^2 \log \left( 1 +
\alpha^2 \left(1-{M\over M_0}\right)^2 \left( \frac{\hat{\varphi}_i -
\hat{\varphi}_j}{M} \right)^2 \right). }
\eear
$$
The final expression, \eq{correct}, is therefore replaced by
$$
S_{tot} \simeq {1 \over 64 \pi^2} \int \sum_{i>j} \left( \hat{W}_i - \hat{W}_j \right)^2
\log \left( \frac{\hat{\varphi}_i -
\hat{\varphi}_j}{\Lambda} \right)^2 ,
$$
$\Lambda$ being the dynamically generated cutoff.

In the $SU(N)$ case, the matrices $F_3, F_8$ and $\{F_i\}_{i \neq 3,8}$
generalise to 
the hermitian $\{H^i\}_{i=1}^r$, where $r = \mbox{rank}(G)$ and to the complex
$\{E^\alpha\}$, $\alpha = 1,\ldots, \mbox{dim}(G) -r$ respectively.

The commutators between $H^i$'s and $E^\alpha$'s are given by the root
equation
$$
\left[ H^i, E^\alpha \right] = \alpha^i E^\alpha, \qquad \qquad 1\leq i
\leq r.
$$

In the case of $SU(N)$, $r \ug N-1$ and $\mbox{dim}(G) -r = N
(N-1)$.

Let us now introduce the hermitian matrices $A^\alpha$ and $B^\alpha$,
defined by
$$
E^\alpha = A^\alpha + i B^\alpha, \qquad  \qquad (E^\alpha)^{*} = E^{-\alpha} =
A^\alpha - i B^\alpha.
$$
The root equation in terms of $A^\alpha$'s and $B^\alpha$'s reads
\be \label{rooteq}
 \left[ H^i, A^\alpha \right] = i \alpha^i B^\alpha, \qquad \qquad  
\left[ H^i, B^\alpha \right] = -i \alpha^i A^\alpha,
\ee
the total number of different $(E^\alpha, E^{-\alpha})$ or $(A^\alpha,
B^\alpha)$ pairs being $\um N (N-1)$.

As in the previous section, we specialise to the diagonal configuration,
$\sum_i H^i \varphi_i$, which in the fundamental representation can be
written as a $N \times N$ diagonal matrix, $\mbox{diag}
\left( \hat{\varphi}_1, \ldots, \hat{\varphi}_{N} \right)$, with $\sum_a
\hat{\varphi}_a = 0$.

In order to evaluate that matrix element in the adjoint representation, we
can read off the relevant structure constants from \eq{rooteq}. Writing
$B^\al$ as $A^{\bara}$ and $ \left[ H^i, A^{\alpha (\bara)} \right] =
i C^{i \al (\bara) \beta} A^\beta$, one has $C^{i \al (\bara) \beta} = 0$
unless $\beta = \bara (\al)$, $C^{i \al \bara} = - C^{i \bara \al} =
\alpha^i$.
 
Thus the non-trivial matrix elements of the Cartan sub-algebra in the
adjoint representation are given by $H^i_{\al \bara} = - H^i_{\bara \al} =
- i \alpha^i$. This means that $\vec{\varphi} \cdot \! \hat{H}$ can be
recast as

\begin{eqnarray} 
\label{matrix}
\vec{\varphi} \cdot \! \hat{H} =
\left( \, 
\begin{tabular}{ccc|rcl|rcl|c}
$0$&&&&&&&&& \\
&$\ddots$& &&&&&&&\\ 
&&$0$&&&&&&& \\ \hline 
&&&&&&&&&\\
&&&&$\vec{\alpha} \cdot \vec{\varphi} \, \sigma_2$&&&&&\\
&&&&&&&&&\\ \hline
&&&&&&&&&\\
&&&&&&&$\vec{\beta} \cdot \vec{\varphi} \, \sigma_2$&&\\
&&&&&&&&&\\ \hline
&&&&&&&&&$\ddots$\\
\end{tabular}
\, \right), \\ 
\underbrace{\hspace{4.7em}}_{r} \hspace{1em}
\underbrace{\hspace{5.05em}}_{\al \bara} \hspace{1.2em}
\underbrace{\hspace{5.05em}}_{\beta \bar{\beta}} \hspace{4.65em} \nonumber \\
\nonumber  
\end{eqnarray}
again with all other elements vanishing.

In order to find out what the values of $\vec{\alpha} \cdot \vec{\varphi}$
are, we can make use of \eq{rooteq} as applied to the fundamental
representation of $SU(N)$: $(\vec{\varphi} \cdot \! \hat{H})_{ab} =
\hat{\varphi}_a \d{ab}$ yields 
$$
\left( \hat{\varphi}_a - \hat{\varphi}_b \right) E^\al_{ab} = \left(\vec{\alpha}
\cdot \vec{\varphi} \right) E^\al_{ab},
$$
\ie\\
i.\phantom{ii} $E^\al_{ab} = 0$ unless the indices $(a,b)$ are such that
$\left( \hat{\varphi}_a - \hat{\varphi}_b \right) = \vec{\alpha} 
\cdot \vec{\varphi}$;\\
ii.\phantom{i} for a given $\vec{\alpha}$, there is at least one pair of
indices $(a,b)$ such that $\left( \hat{\varphi}_a - \hat{\varphi}_b \right)
= \vec{\alpha} \cdot \vec{\varphi}$;\\
iii. for different roots $\vec{\alpha}, \vec{\beta}$, excluding the
case $\vec{\alpha} = -\vec{\beta}$, the pairs of indices in ii. are\\
\phantom{iii.} different to each other.

As a consequence, when $\vec{\al}$ goes over all the 
$(E^\al,E^{-\al})$ pairs, the corresponding $\left( \hat{\varphi}_a -
\hat{\varphi}_b \right)$ takes all the different index pairs. Since the
number of different $(E^\al,E^{-\al})$ pairs equals that of $(a,b)$
combinations, one can rewrite \eq{matrix} as
$$
\vec{\varphi} \cdot \! \hat{H} = (0)^r \otimes \prod_\al \left( \vec{\alpha} 
\cdot \vec{\varphi} \right) \sigma_2 = (0)^r \otimes \prod_{a>b} \left(
\hat{\varphi}_a - \hat{\varphi}_b \right) \, \sigma_2. 
$$

These results can be used to evaluate the trace of the logarithmic term in
\eq{kin2} once $\hat{F}$ has been replaced by $\hat{H}$.
Making use of \eq{logexp} and 
writing $W^2 = (0)^r \otimes \prod_{a>b} \left(
\hat{W}_a - \hat{W}_b \right)^2 \!\! \one$ one gets
\be \label{sanomsun}
\bear
{\ds 
\sum_{a>b} \left( \hat{W}_a - \hat{W}_b \right)^2 \!
\log \left( 1 +
\gamma^2 \! \left(1-{M\over M_0}\right)^2 \! \left( \frac{\hat{\varphi}_a -
\hat{\varphi}_b}{M} \right)^2 \right)= }\nonumber\\[0.35cm]
{\ds = \sum_{\al} \left( \vec{\al} \cdot \vec{W} \right)^2
\log \left( 1 +
\gamma^2 \left(1-{M\over M_0}\right)^2 \left( \frac{\vec{\al} \cdot 
\vec{\varphi}}{M} \right)^2 \right),}
\eear
\ee
where the constant $\alpha$ in the definition of $\hat{A}(\varphi)$
[cf. \eq{A}] has been replaced by $\gamma$ to avoid confusion with
$\vec{\alpha}$, defined by the root equation.  

Hence the $W^2$-term in the effective action $S_M$ is
$$
\bear
{\ds \frac{1}{16} \int \left( \frac{1}{g_0^2} + \frac{N}{4 \pi^2} \log
\left( {M\over M_0} \right) \right) W^a_{\alpha} W^{a \alpha}  
- \frac{1}{16} \frac{1}{4 \pi^2} \sum_{a>b} \int
\left( \hat{W}_a - \hat{W}_b \right)^2 \times }\\[0.35cm]
{\ds \times \left( 1 +
\gamma^2 \! \left(1-{M\over M_0}\right)^2 \! \left( \frac{\hat{\varphi}_a -
\hat{\varphi}_b}{M} \right)^2 \right). } 
\eear
$$

\section{Conclusion and comments}

The results given in the previous section represent the first application
of the method proposed in~\cite{noi} beyond those already given
in~\cite{mu1,mu2}, \ie the exact NSVZ beta-functions.

The local potential approximation adopted here is of course expected to
give only a rough estimate of the holomorphic part of the effective action,
$S_M$. Moreover, it explicitly breaks \ntwo \ supersymmetry since one
cannot treat the non-holomorphic kinematical terms for chiral
superfields on the same footing as the kinematical term for the vector field.  

Nevertheless, \eq{correct} seems to indicate that the ERG equation applied
to the mass deformed \nfo \ SYM retains, at least, some correct information
about the low energy behaviour of \ntwo \ SYM. The same method can be
applied to other models such as $\N \ug 1^*$.

In order to see whether non-perturbative corrections (instanton
contributions)~\cite{Se,SW} too can be obtained in this way or not, one
needs a more accurate estimate of the solution of Polchinski's equation. 

Actually, except in the infinite $\varphi \over \Lambda$ limit, the heavy
field dependent residual potential in \eq{smlpa} produces also some
corrections of order $({\Lambda \over \varphi})^{2n}, n \in \nn$, but they
are unlikely to be instanton effects, as they should be of order $({\Lambda
\over \varphi})^{4n}, n \in \nn$ being the instanton number.

It is also conceivable that we should consider corrections to the Konishi
anomaly itself beyond what has been computed in~\cite{Kon}, that is the
anomalous Jacobian determinant in presence of an external vector field,
just like the chiral anomaly. This would be the ``quantum anomaly''
discussed in~\cite{Pronin}, but at present we do not understand the
consistent way of calculating such a correction.

\acknowledgments{The authors wish to thank their colleagues for useful
discussions and encouragement, in particular M. Bianchi, N. Evans,
K. Konishi, S. Kovacs, T.R. Morris, G.C. Rossi and Y. Stanev.

KY wishes to thank T. Eguchi, K. Fujikawa, H. Kawai and in particular
Y. Nomura for constructive comments. He also wishes to thank Prof. Fujikawa
for the hospitality at the Department of Physics, University of Tokyo,
where part of the work has been carried out.}


\end{document}